\documentclass[11pt]{iopart}

\newcommand{\MP}{M_{\rm P}}
\newcommand{\gsim}{\mbox{\raisebox{-1.ex}{$\stackrel{\textstyle>}{\textstyle\sim}$}}}
\renewcommand{\lsim}{\mbox{\raisebox{-1.ex}{$\stackrel{\textstyle<}{\textstyle\sim}$}}}

\begin{document}
\title{Gravitational Wave Constraints on DBI Inflation}
\date{\today}
\author{James E. Lidsey and Ian Huston}
\address{Astronomy Unit, School of Mathematical Sciences\\
  Queen Mary, University of London\\
  Mile End Road, London E1 4NS\\
  United Kingdom}
\eads{\mailto{J.E.Lidsey@qmul.ac.uk}, \mailto{I.Huston@qmul.ac.uk}}
\submitto{JCAP}
\begin{abstract}
An upper bound on the amplitude of the primordial gravitational wave spectrum
generated during ultra-violet DBI inflation is derived. 
The bound is insensitive to the form of the inflaton
potential and the warp factor of the compactified 
dimensions and can be expressed entirely in terms of 
observational parameters once the volume of the five-dimensional 
sub-manifold of the throat has been specified. For standard type IIB 
compactification schemes, the bound predicts 
undetectably small tensor perturbations 
with a tensor-scalar ratio $r < 10^{-7}$. 
This is incompatible with a corresponding lower limit of 
$r > 0.1 (1-n_s )$,  which applies to any model 
that generates a red spectral index $n_s <1$ and a potentially 
detectable non-Gaussianity in the curvature perturbation. 
Possible ways of evading these bounds in more 
general DBI-type scenarios are discussed and a 
multiple-brane model is investigated as a specific example. 

\vspace{3mm}
\begin{flushleft}
  \textbf{Keywords}:
  Inflation,
  Physics of the early universe.
\end{flushleft}
\end{abstract}
\maketitle

\section{Introduction}

\label{sec:intro}

The inflationary scenario provides the 
theoretical framework for the early history 
of the universe. It has now been successfully tested by observations, 
including the third year data from the Wilkinson Microwave Anisotropy 
Probe (WMAP3) \cite{spergel}. Despite this success, however, the high energy 
physics that resulted in a phase of accelerated expansion is still 
not well understood. String/M-theory unifies the fundamental interactions 
including gravity and it is natural and important to 
develop inflationary models within string theory and to confront them with 
cosmological observations. 
 
One class of string-theoretic models that has received 
considerable attention is D-brane inflation
\cite{brane1,brane2,brane3,brane4,brane5,brane6,brane7,brane8,brane9,brane10,brane11,brane12,brane13,brane14,brane15,brane16,brane17,Brodie:2003qv,Vikman:2006hk,Mukhanov:2005bu,Kallosh:2007wm,brane18,brane19,brane20,brane21}. 
(For recent reviews, see, e.g., \cite{tyereview,cline}). 
Of particular interest 
is the DBI inflationary scenario \cite{brane6,brane11}. 
This is based on the compactification of type IIB string theory on a 
Calabi-Yau (CY) three-fold, where the form-field fluxes generate locally
warped regions known as `throats'.  The propagation of a 
${\rm D3}$-brane in such a region can drive inflation, where the inflaton 
field is identified with the radial position of the brane 
along the throat. Since this is an open string mode, the field 
equation for the inflaton is determined by a Dirac-Born-Infeld (DBI) action. 

The purpose of the present work is to 
explore the observational consequences of DBI inflation. 
In general, primordial gravitational wave fluctuations
and non-Gaussian statistics in the curvature perturbation provide 
two powerful discriminants of inflationary scenarios. 
The nature of the DBI action is such that the sound 
speed of fluctuations in the inflaton can be much less than the speed of 
light. This induces a large and potentially detectable non-Gaussian 
signal in the density perturbations \cite{brane6,brane11,lidser3,chenetal}. 

The gravitational wave background generated in DBI 
inflation was recently investigated by Baumann and McAllister (BM) 
\cite{bmpaper}. By exploiting a relationship due originally 
to Lyth \cite{lyth}, these authors derived a field-theoretic upper limit 
to the tensor amplitude and concluded that 
rather stringent conditions would need to be satisfied for these 
perturbations to be detectable.      
Moreover, the special case of 
DBI inflation driven by a quadratic potential is incompatible with the WMAP3 
data when this constraint is imposed \cite{bean}.  

Our aim is to derive observational constraints on DBI inflation that are 
insensitive to the details of the throat geometry and the inflaton potential. 
In general, there are two realizations of the scenario, 
which are referred to as the ultra-violet (UV) and infra-red (IR) 
versions, respectively. These are characterized by whether the brane is 
moving towards or away from the tip of the throat. 
We focus initially on the UV scenario 
and derive an upper bound on 
the gravitational wave amplitude in terms of observable 
parameters. This limit arises by considering 
the variation of the inflaton field during the era when 
observable scales cross the Hubble radius, and 
we find in general that the tensor-scalar ratio must satisfy 
$r \, \lsim \, 10^{-7}$. This 
is below the projected sensitivity of future Cosmic Microwave Background (CMB) polarization 
experiments \cite{clover,vpj}. On the other hand, the WMAP3 data 
favours a red perturbation spectrum with 
$n_s<1$ when the tensor modes are negligible and 
the scalar spectral index is effectively constant \cite{spergel}. 
For models which generate such a spectrum, 
we identify a corresponding lower limit on the 
tensor modes such that $r \, \gsim \, 0.1 (1-n_s)$. 
This is incompatible with the upper bound 
on $r$ when $1-n_s \simeq 0.05$, as inferred
by the observations. 

Hence, a reconciliation between theory and observation 
requires either a relaxation of the upper limit on $r$ or a blue 
spectral index $(n_s >1)$. The DBI scenario would need 
to be generalized in a suitable way for the upper bound on $r$
to be weakened. We identify necessary conditions that a 
generalized action must satisfy for the Baumann-McAllister
 constraint to be relaxed. 
We then show that such conditions can be realized in a recently 
proposed IR version of DBI inflation driven
by multiple coincident branes \cite{thomasward}. 

The paper is organized as follows. We briefly review DBI inflation 
in Section \ref{sec:dbiinflation}. In Sections \ref{sec:upper} and  
\ref{sec:lower}, we derive the respective 
limits on the amplitude of the tensor perturbations.  
We determine how the BM bound may be relaxed in generalized DBI models in 
Section \ref{sec:satisfy} and discuss multi-brane 
IR scenarios in Section \ref{sec:multiple}. Finally, we conclude  
in Section \ref{Conclusion}.   
Units are chosen such that $\MP \equiv (8\pi G )^{-1/2}= 2 
.4 \times 10^{18}\, {\rm GeV}$ 
defines the reduced Planck mass and $c=\hbar =1$. 

\section{DBI inflation} 

\label{sec:dbiinflation}

Flux compactification of type IIB string theory to four dimensions 
results in a warped geometry, where the six-dimensional CY  
manifold contains one or more throats (see \cite{grana,douglas} for reviews). 
The metric inside a throat has the generic form
\begin{equation}
\label{conemetric}
ds_{10}^2= h^2 ( \rho) ds_4^2 + h^{-2} (\rho ) 
\left( d\rho^2 +\rho^2 ds_{X_5}^2 \right) \,,
\end{equation} 
where the warp factor $h (\rho)$ is a function of the 
radial coordinate $\rho$ along the throat and $X_5$
is a Sasaki-Einstein five-manifold. 
In many cases, the ten-dimensional metric (\ref{conemetric}) can be 
approximated locally by the geometry $AdS_5 \times X_5$, where the 
warp factor is given by $h=\rho /L$ and 
the radius of curvature of the $AdS_5$ space is defined by 
\begin{equation}
L^4 \equiv \frac{4\pi^4 g_s N}{{\rm Vol} (X_5) m_s^4} \,,
\end{equation}
such that $\mathrm{Vol}(X_5)$ is the dimensionless volume of 
$X_5$ with unit radius, $N$ is the ${\rm D3}$  charge of the throat, 
$g_s$ is the string coupling and $m_s$ is the string mass scale.
In the Klebanov-Strassler (KS) background \cite{ks}, the throat 
is a warped deformed conifold and 
corresponds to a cone over the manifold 
$X_5 = T^{1,1}= {\rm SU(2)} \otimes {\rm SU(2)}/{\rm U(1)}$
in the UV limit ($\rho \rightarrow \infty$). This has   
a volume $\mathrm{Vol} (T^{1,1}) = 16\pi^3/27$ and  topology
$S^2\times S^3$, where the $S^2$ is fibred over the $S^3$.
The conical singularity at the tip of the throat 
is smoothed out by an 
$S^3$ `cap' due to the wrapping of the fluxes along the cycles of the conifold
\cite{ks,kt} and the warp factor asymptotes to 
a constant value in this region.   
 
In general, the low-energy world-volume dynamics
of a probe ${\rm D3}$-brane in a warped throat is determined 
by an effective, four-dimensional DBI action. 
The inflaton field is related to the radial 
position of the brane by 
$\phi \equiv \sqrt{T_3} \rho$, where $T_3 =m_s^4/[(2\pi )^3 g_s ]$ 
is the brane tension. The action is then given by 
\begin{eqnarray}
\label{DBIaction}
S=\int  d^4x \sqrt{|g|} \left[ \frac{\MP^2}{2} R 
+ P (\phi , X) \right] \\
\label{defP}
P( \phi ,X) = - T (\phi)  \sqrt{1 - 2T^{-1} (\phi ) X}
+T (\phi)  -V(\phi)  \,,
\end{eqnarray}
where $R$ is the Ricci curvature scalar,
$X \equiv - \frac{1}{2} g^{\mu\nu} \nabla_{\mu} \phi \nabla_{\nu} \phi$
is the kinetic energy of the inflaton, $V(\phi )$ denotes 
the field's interaction 
potential and $T(\phi ) = T_3 h^4 (\phi )$
defines the warped brane tension. We refer to $P(\phi , X)$ as the 
`kinetic function' for the inflaton. 
 
We consider a spatially flat and isotropic cosmology 
sourced by a homogeneous scalar field. 
In this case, the Friedmann equations for a monotonically 
varying inflaton can be expressed in the form \cite{brane6} 
\begin{eqnarray}
\label{Friedmann}
3 \MP^2 H^2(\phi ) = V(\phi ) -T(\phi ) 
\left[ 1- \sqrt{1+4\MP^4T^{-1} H'^2} \right] \\
\label{useful}
\dot{\phi} = - \frac{2\MP^2H'}{\sqrt{1+4\MP^4 T^{-1} H'^2}} \, ,
\end{eqnarray}
where $H=H(\phi )$ represents the Hubble parameter
as a function of the field and a prime denotes $d/d\phi$. 

An important consequence of the non-trivial kinetic structure 
of the DBI action is that the sound speed of fluctuations in the inflaton 
differs from unity:   
\begin{equation}
\label{speedofsound}
c_s = \frac{1}{P_{,X}} = \sqrt{1 -2T^{-1}X}  \,,
\end{equation}
where a subscripted comma denotes partial differentiation. 
Furthermore, the condition that the sound speed be real 
imposes an upper bound on the kinetic energy 
of the inflaton, $\dot{\phi}^2 < T(\phi)$, which 
is independent of the steepness of the potential.
The motion of the brane is said to be `relativistic' when this bound is 
close to saturation.
 
We now define the epoch that is directly 
accessible to cosmological observations as `observable inflation'. 
Depending on the reheating 
temperature, this occurred some 30 to 60 e-foldings before the end of 
inflation. The inflationary dynamics during this phase can  
be quantified in terms of three parameters: 
\begin{eqnarray}
\label{defepsilon}
\epsilon \equiv -\frac{\dot{H}}{H^2}
= \frac{XP_{,X}}{\MP^2H^2} 
= \frac{2\MP^2}{\gamma} \left( \frac{H'}{H} \right)^2 \\
\label{defeta}
\eta \equiv  \frac{2\MP^2}{\gamma}\frac{H''}{H} \\
\label{defs}
s \equiv \frac{\dot{c_s}}{c_sH} 
= \frac{2\MP^2}{\gamma} \frac{H'}{H}\frac{\gamma'}{\gamma}  \,,
\end{eqnarray}
where $\gamma \equiv 1/c_s$. 
We will assume that the `slow-roll' conditions 
$\{ \epsilon, |\eta | , |s | \}  \ll 1$ applied during observable inflation. 
In this regime, the amplitudes and spectral indices of the two-point functions 
for the scalar and tensor perturbations are given by \cite{gm}
\begin{eqnarray}
\label{amplitudes}
P_S^2= \frac{1}{8 \pi^2 \MP^2} \frac{H^2}{c_s \epsilon}
= \frac{H^4}{4\pi^2\dot{\phi}^2}
\\
P_T^2 = \frac{2}{\pi^2} \frac{H^2}{\MP^2} 
\\
\label{indices}
1-n_s = 4 \epsilon -2\eta  +2s 
\\
 n_t = -2\epsilon  
\end{eqnarray}
respectively, where the quantities on the right-hand sides are evaluated 
when the scale with comoving wavenumber $k=aH\gamma$ crossed 
the Hubble radius during inflation.  
The tensor-scalar ratio, $r \equiv P_T^2/P_S^2$, is directly related to 
the tensor spectral index by \cite{gm}
\begin{equation}
\label{consistency}
r= -8c_sn_t = 16c_s \epsilon\,.
\end{equation}
Hence, a sound speed different to unity leads to a violation of the 
standard inflationary consistency equation, which might be 
detectable in the foreseeable future \cite{lidser1,lidser2}. 
Recent observations of the CMB 
indicate that $P^2_S=2.5 \times 10^{-9}$ and $r < 0.55$ \cite{spergel}. 
The best-fit value for a constant spectral index is 
$n_s = 0.987^{+0.019}_{-0.037}$ if $r\ne 0$ is assumed as a prior. 
This is strengthened to 
$n_s = 0.948^{+0.015}_{-0.018}$ for a prior of $r = 0$ \cite{spergel}. 

A further important consequence of a small sound speed is that departures  
from purely Gaussian statistics may be large 
\cite{brane6,brane11,lidser3,chenetal}. It is conventional 
to quantify such deviations by expressing the non-Gaussian curvature 
perturbation ${\cal{R}}$ in the form 
${\cal{R}} ={\cal{R}}_G+f_{NL}
({\cal{R}}^2_G -\langle {\cal{R}}^2_G \rangle)$, where 
${\cal{R}}_G$ represents the Gaussian contribution, 
the quadratic piece is a convolution and $f_{NL}$ defines 
the `non-linearity' parameter \cite{maldacena}. 
In general, this parameter is a function of the three momenta which 
form a triangle in Fourier space. However, in the limit where 
these momenta have equal magnitude, the non-linearity parameter 
is given to leading-order by \cite{chenetal,lidser2}  
\begin{equation}
\label{fnlcs}
f_{NL} = \frac{1}{3} \left( \frac{1}{c_s^2} -1 \right) \,.
\end{equation}
Currently, WMAP3 constrains this parameter  
to be ${| f_{NL} |} \, {\lsim} \, {300}$ \cite{spergel,crim}.  

Finally, Eqs. (\ref{speedofsound}), (\ref{amplitudes}) and (\ref{fnlcs}) 
may be combined to constrain the warped brane tension 
during observable inflation: 
\begin{equation}
\label{observewarp}
\frac{T (\phi_*)}{\MP^4}  = 
\frac{\pi^2}{16} r^2P_S^2 \left( 1+\frac{1}{3f_{NL}} \right) \,,
\end{equation}
where a subscript `$*$' denotes that the quantity is to be evaluated 
at that epoch. 

In the following, we first consider the UV version of DBI inflation,
where the brane moves relativistically 
towards the tip of the throat. We will assume implicitly 
that the sound speed is sufficiently small to generate a non-Gaussianity in 
excess of $f_{NL} \, {\gsim} \, 5$, since this is the projected 
sensitivity of the Planck satellite \cite{planck}. Furthermore, we consider   
an arbitrary warp factor and inflaton potential, 
subject only to the condition that a sufficiently long phase of 
quasi-exponential expansion can be maintained to solve the horizon problem of
the hot big bang model.

\section{An Upper Bound on the Primordial Gravitational Waves}

\label{sec:upper}

Recently, Baumann and McAllister (BM) \cite{bmpaper} 
derived a field-theoretic upper bound on the tensor-scalar ratio 
by noting that the four-dimensional Planck mass is related 
to the volume of the compactified CY manifold, $V_6$, such that 
$\MP^2=V_6 \kappa_{10}^{-2}$, where $\kappa_{10}^2 \equiv 
\frac{1}{2} (2\pi )^7 g_s^2 m_s^{-8}$. In general, the compactified volume 
is comprised of bulk and throat contributions, 
$V_6 = V_{6,{\rm bulk}}+V_{6,{\rm throat}}$, where the latter is 
given by
\begin{equation}
\label{throatvolume}
V_{6,\mathrm{throat}} = \mathrm{Vol}(X_5)  
\int_0^{\rho_{UV}} d\rho \frac{\rho^5}{h^4(\rho )} \,,
\end{equation}
and $\rho_{UV}$ denotes the radial coordinate at 
the edge of the throat (defined as the region 
where $h (\rho_{UV})$ is of order unity). 
If one assumes that the bulk volume is 
non-negligible relative to 
that of the throat (i.e. $V_{6,\mathrm{throat}} < V_{6}$), 
it follows that $\MP^2> V_{6,\mathrm{throat}}\kappa_{10}^{-2}$. 
For a warped $AdS_5 \times X_5$ geometry, this leads to an 
upper limit on the total variation of the inflaton field in 
the throat region:
\begin{equation}
\label{BMbound}
\frac{\phi_{UV}}{\MP}   < \frac{2}{\sqrt{N}} \,.
\end{equation}

Condition (\ref{BMbound}) may be converted into a 
corresponding limit on the tensor-scalar ratio by noting from 
the definition (\ref{defepsilon})
that $\dot{\phi}^2 /\MP^2 = 2\epsilon H^2/P_{,X}$.
This implies that the variation of the inflaton field 
is given by the Lyth bound 
\cite{lyth,bmpaper}
\begin{equation}
\label{rtheory}
\frac{1}{\MP^2} \left( \frac{d\phi}{d {\cal{N}}} \right)^2 =
\frac{r}{8} \,,
\end{equation}
where ${\cal{N}} \equiv \int dt H$. 
Since $\phi_* < \phi_{UV}$, this 
results in an upper bound on the observable tensor-scalar ratio
\cite{bmpaper}: 
\begin{equation}
\label{BMboundr}
r_*  < \frac{32}{N ({\cal{N}}_{\mathrm{eff}})^2} \,,
\end{equation}
where ${\cal{N}}_{\mathrm{eff}} \equiv 
\int^{{\cal{N}}_{\mathrm{end}}}_0 d {\cal{N}} (r/r_*)^{1/2}$ 
is a model-dependent parameter that quantifies 
how $r$ varies during the final stages of inflation.  

Typically, one expects $30 \, \lsim \, {{\cal{N}}_{\mathrm{eff}}} 
\, \lsim \, 60$, 
although smaller values may be possible if the slow-roll conditions are 
violated after observable scales have crossed the horizon. 
Furthermore, $N \gg 1$ is necessary 
for backreaction effects to be negligible. Hence, the 
constraint (\ref{BMboundr}) 
imposes a strong restriction on DBI inflationary models. 
On the other hand, the numerical value 
of ${\cal{N}}_{\mathrm{eff}}$ is uncertain.  
Our aim here is to focus on the range of values covered by the 
inflaton field during the observable stages of inflation. 
This will result in a constraint on the tensor modes that 
can be expressed in terms of observable parameters.  

To proceed, we denote the change in the value of the inflaton field over 
observable scales by 
$\Delta  \phi _{*} = \sqrt{T_3} \Delta \rho_{*}$. 
Since the brane moves towards the tip of the throat in 
UV DBI inflation, it follows that $\rho_{*} > \rho_{end} >0$, which 
implies that  
\begin{equation}
\label{importantbound}
\rho_{*} > |\Delta \rho _{*}| \,.
\end{equation}
This change in the inflaton value will correspond 
to a fraction of the throat volume, 
$| \Delta V _{6,*}|  < {V_{6,\mathrm{throat}}} \, \lsim \, {V_6} $,
where equality in the second limit arises if
the bulk volume is negligible. Hence, 
$| \Delta \phi_* |$ is bounded such that  
\begin{equation}
\label{halfwayconstraint}
\left( \frac{\Delta \phi}{\MP} \right)^2_{*} < 
\frac{T_3 \kappa_{10}^2 (\Delta \rho_{*})^2}{|\Delta V_{6,*}|} \,.
\end{equation} 

The observations of the CMB 
that directly constrain the primordial tensor perturbations only 
cover multipole values in the range $2 \le l \, \lsim \, 100$. 
This is equivalent to no more than ${\Delta {\cal{N}}_{*}} \simeq {4}$ 
e-foldings of inflationary expansion and, in general,   
corresponds to a narrow range of inflaton values. 
To a first approximation, therefore, the fraction of the throat volume 
(\ref{throatvolume}) that is accessible to cosmological 
observation can be estimated to be 
\begin{equation}
\label{trapezium}
| \Delta V_{6,*} | \simeq \mathrm{Vol}(X_5) 
\frac{|\Delta \rho_*| \rho^5_{*}}{h^{4}_{*}} \,.
\end{equation}
Combining the inequality (\ref{importantbound}) with Eq. (\ref{trapezium}) 
then implies that 
\begin{equation}
\label{trapeziumlimit}
|\Delta V _{6,*}| > \mathrm{Vol}(X_5) 
\frac{(\Delta \rho_* )^6}{h^{4}_*}  \,,
\end{equation}
and substituting the CMB normalization (\ref{observewarp}) and 
condition (\ref{trapeziumlimit}) into the bound (\ref{halfwayconstraint}) 
yields the upper limit   
\begin{equation}
\label{boundpower6}
\left( \frac{\Delta \phi}{\MP} \right)^6_{*} 
< \frac{\pi^3}{16\mathrm{Vol}(X_5)} r^2 P_S^2 
\left( 1+ \frac{1}{3f_{NL}} \right)  \,.
\end{equation}
Hence, employing the Lyth bound in the form $(\Delta \phi_* / \MP )^2 \simeq 
r (\Delta {\cal{N}}_*)^2 /8$  
results in a very general upper limit on the tensor-scalar ratio: 
\begin{equation}
\label{generalbound}
r_{*} < \frac{32 \pi^3}{(\Delta {\cal{N}}_*)^6 \mathrm{Vol}(X_5)} 
P_S^2 \left( 1+ \frac{1}{3f_{NL}} \right) \,.
\end{equation}

Condition (\ref{generalbound}) is 
only weakly dependent on the level of non-Gaussianity 
when $f_{NL} > 5$ and we may therefore neglect the 
factor involving this parameter. Substituting the WMAP3 normalization 
$P_S^2 = 2.5 \times 10^{-9}$ then implies that 
\begin{equation}
\label{upperbound}
r_{*} < \frac{2.5\times 10^{-6}}{( \Delta {\cal{N}}_*)^6 \mathrm{Vol}(X_5)} \,.
\end{equation}
Furthermore, the most optimistic 
estimate for the minimum number of e-foldings that could be 
probed by observation is $\Delta {\cal{N}}_{*} \simeq 1$, whereas
a generic compactification arises when 
the volume of the Einstein five-manifold is $\mathrm{Vol}(X_5) 
\simeq {\cal{O}} (\pi^3)$ \cite{ks}. This yields a model-independent 
upper bound on the tensor-scalar ratio for standard UV DBI inflation:   
\begin{equation}
\label{standardbound}
r_* < 10^{-7} \,.
\end{equation}
This is significantly below the sensitivity 
of future CMB polarization experiments, which will measure 
${r} \, {\gsim} \, 10^{-4}$ \cite{clover,vpj}. If CMB  
observations are able to span the full range of e-foldings such that
$\Delta {\cal{N}}_* \simeq 4$, this constraint is strengthened to 
${r_*} \, {\lsim} \, {2 \times 10^{-11}}$.

Before concluding this section, we should remark that 
the estimate (\ref{trapezium}) was derived under the assumption  
that the integrand in Eq. (\ref{throatvolume})  
is constant. This inevitably introduces errors into the bound
(\ref{generalbound}). However, the two limiting cases of interest 
in KS-type geometries arise 
when the warp factor scales either as $h \propto \rho$
or as $h \simeq {\rm constant}$ \cite{ks,kt}. In both cases
the integral (\ref{throatvolume}) can be performed analytically. 
Indeed, if we specify $h \propto \rho^{\alpha}$ for some constant $\alpha$,  
evaluate the integral between $\rho_{*}$ 
to $\rho_{*}+\Delta \rho_{*}$, and expand to second-order in a 
Taylor series, we find that  
\begin{equation}
\label{limits}
\Delta V_{6,*} \simeq \mathrm{Vol}(X_5) \frac{\rho^5_{*}}{h^{4} 
(\rho_{*} )}(\Delta \rho_*) 
\left[ 1 +\frac{(5-4 \alpha )}{2} 
\frac{(\Delta \rho_*)}{\rho_{*}} \right]  \,.
\end{equation}
This implies that the error in Eq. (\ref{trapezium}) 
is no greater than 
about $3 (\Delta \rho_* / \rho_*)$ if  
$0 \le \alpha \le 1$. More generally, it follows that a similar
error will arise for {\em any} warp factor 
$h \propto \rho^{\alpha (\rho )}$, where the function 
$\alpha (\rho)$ satisfies $0 \le \alpha (\rho ) \le 1$ 
over observable scales. 
We conclude, therefore, that Eq. (\ref{trapezium}) 
provides a sufficiently good estimate of the volume element  
for a generic warp factor\footnote{As we shall see in the following Section, 
even an order of magnitude error will make little 
difference to our final conclusions.}.  

\section{A Lower Bound on the Primordial Gravitational Waves} 

\label{sec:lower}

The analysis of the previous Section 
indicates that standard versions of UV DBI inflation generate a 
tensor spectrum that is unobservably 
small. Hence, $r=0$ can be assumed as a prior 
when discussing the WMAP3 data. However, in this case the data 
excludes a scale-invariant density spectrum at the $3 \sigma$ level
when the running in the spectral index, $\alpha_s \equiv dn_s/d\ln k$, 
is negligible \cite{spergel}.  
Furthermore, a blue spectral index 
is only marginally consistent with the data when $r\ne 0$ and $\alpha_s=0$. 
(The inferred upper limit is $n_s < 1.006$).
Although the results from WMAP3 do allow for a blue spectrum if there is 
significant negative running in the spectral index, we will 
focus in this Section 
on models that generate a red spectral index $n_s<1$, since these seem 
to be preferred by the current data.   

In general, the spectral index may be related to the tensor-scalar ratio. 
After differentiating Eq. (\ref{speedofsound}) 
with respect to cosmic time, and employing Eqs. (\ref{useful}) 
and (\ref{indices}), we find that 
\begin{equation}
\label{constraint}
1-n_s = 4 \epsilon +\frac{2s}{1-\gamma^2} \mp 
\frac{2\MP^2}{\gamma} \frac{T'|H'|}{TH}  \,,
\end{equation}
where the minus (plus) sign corresponds to 
a brane moving down (up) the warped throat. 
The second term in Eq. (\ref{constraint}) 
can be converted into observable parameters
by defining the `tilt' of the non-linearity parameter, 
$n_{NL} \equiv d \ln f_{NL}/d\ln k$ \cite{brane14}. 
This implies that $s= - 3 f_{NL} n_{NL} /[2(1+3f_{NL})]$ and     
substitution of Eqs. (\ref{indices})--(\ref{fnlcs}) 
into Eq. (\ref{constraint}) then yields  
\begin{equation}
\label{obscon1}
1-n_s = \frac{r}{4} \sqrt{1+3f_{NL}} + \frac{n_{NL}}{1+3f_{NL}}
\mp \sqrt{\frac{r}{8}} \left( \frac{T'}{T} \MP \right)_*  \,.
\end{equation}

In \cite{lidser2}, brane inflation near the tip of a KS-type 
throat was considered, where the warped brane tension asymptotes to a 
constant value. In this regime, Eq. (\ref{obscon1}) reduces to 
the condition $r \simeq 2.3 (1-n_s)/\sqrt{f_{NL}}$ when $f_{NL}$ is 
sufficiently large to be detectable by Planck, 
i.e., $|f_{NL}| > 5$. It then follows from the  
WMAP3 best-fit value $n_s \simeq 0.987$ and upper limit 
$|f_{NL}| < 300$ \cite{crim} that the gravitational wave amplitude 
is bounded both from above and below such that $0.001 \, \lsim \, r 
\, \lsim \, 0.01$. These bounds follow from 
current WMAP3 limits on the spectral index and the 
non-linearity parameter, but do not take into account the 
field-theoretic upper bound that must be imposed 
on the variation of the inflaton field during inflation. 

More generally, in UV DBI inflation where the brane moves towards the 
tip of the throat, it is reasonable to assume 
that the warp factor decreases monotonically 
with the radial coordinate over the observable range of inflaton values, 
i.e., that $dh/d \rho \ge 0$. This condition is satisfied for  
$AdS_5 \times X_5$ compactifications and KS-type solutions. 
Consequently, the third term in 
Eq. (\ref{obscon1}) will be semi-negative definite, 
which implies that 
\begin{equation}
\label{halfbound}
\frac{r}{4} \sqrt{1+3f_{NL}} + \frac{n_{NL}}{1+3f_{NL}} 
> 1-n_s \,.
\end{equation}

Condition (\ref{halfbound}) is a consistency relation on UV DBI 
inflation in terms of observable parameters and it 
may be combined with the upper bound 
(\ref{generalbound}) to confront the scenario with observations.
Firstly, let us assume that the tensor-scalar ratio is negligible. 
The WMAP3 data implies that $1-n_s > 0.037$, and this is only 
compatible with condition (\ref{halfbound}) if 
\begin{equation}
\label{anotherbound}
n_{NL} \simeq -2s > 3 (1-n_s ) f_{NL} > 0.1 f_{NL} \,.
\end{equation}
However, when $f_{NL} \gg 1$, this would violate the slow-roll conditions
that must be satisfied for a consistent 
derivation of the perturbation spectra 
(\ref{amplitudes}). For example, the conservative 
bound $|s| < 0.1$ with $1-n_s \simeq 0.05$ is violated if  
$f_{NL} > {\cal{O}} (5)$. 

In view of this, let us consider the case where the tensor 
perturbations are non-negligible. 
The magnitude of the second term in condition (\ref{halfbound}) 
is suppressed by a factor of $f_{NL}^{3/2} \gg 1$ 
relative to the first. This is expected to 
be a significant effect in DBI inflation. Consequently, 
by saturating the WMAP3 limit $|f_{NL}| < 300$ \cite{crim}, we arrive at 
a lower bound on the tensor-scalar ratio which applies   
to any model for which the ratio $(n_{NL}/f_{NL})$ is 
negligible:
\begin{equation}
\label{lowerbound}
r_* >  \frac{4(1-n_s)}{\sqrt{3f_{NL}}} > \frac{1-n_s}{8} \,.
\end{equation}
This requires $r > 0.002$ for the WMAP3 best-fit value 
$1-n_s \simeq 0.013$, which is incompatible with the upper limit 
(\ref{standardbound}). 

In general, therefore, it is difficult to simultaneously satisfy 
the bounds on $r$ with the WMAP3 data
in standard UV DBI inflation. There is a 
small observational window where a blue spectrum is consistent 
with the data, in which case the lower limit 
(\ref{lowerbound}) does not apply. 
However, if the tensor modes are negligible,
as implied by the inequality (\ref{standardbound}), the 
data strongly favours a red spectral index with $n_s < 0.963$,
and this violates the condition (\ref{lowerbound}). A significant 
detection of a red spectral index requires either a 
violation of the slow-roll conditions or a sufficiently 
small value for the volume of $X_5$. 
In particular, combining the limits
(\ref{upperbound}) and (\ref{lowerbound}) results in the condition 
\begin{equation}
\label{upperboundvol}
\mathrm{Vol}(X_5) < \frac{2 \times 10^{-5}}{(1-n_s) 
(\Delta {\cal{N}}_{*} )^6}  \,,
\end{equation}
and we find that $\mathrm{Vol}(X_5) \, \lsim \, 10^{-7}$ 
for typical values $1-n_s \simeq 0.05$ and 
$\Delta {\cal{N}}_* \simeq 4$. This 
is comparable to the limit on the volume derived for the special case of a 
quadratic inflaton potential \cite{bmpaper}.  
As noted in \cite{bmpaper,bean}, condition 
(\ref{upperboundvol}) may be achieved 
if $X_5$ corresponds to a $Y^{p,q}$ space, 
which has arbitrarily small volume in the limit  
$q =1$ and $p \rightarrow \infty$ \cite{gauntlett}. 
Small volumes could also be realized 
by orbifolding the $S^2$ symmetry of a KS-type throat. 

On the other hand, 
our upper limit (\ref{upperbound}) on the gravitational waves 
follows as a consequence of assuming 
the constraint (\ref{importantbound}). This  
could be violated in IR versions of the scenario, where
observable scales crossed the Hubble radius when the 
brane was near the tip of the throat and $\phi \ll \MP$
\cite{brane12,brane14}. 
Nonetheless, we emphasize that the upper bound (\ref{upperbound})
on the tensor modes 
will also apply to any IR DBI model for which 
$|\Delta \phi_* | < \phi_*$.  
In view of the above discussion, therefore, 
we will proceed in the following Section
to discuss a framework for generalizing the DBI scenario so 
that the constraints on the tensor modes can be satisfied. 

\section{Relaxing the Baumann-McAllister Bound}

\label{sec:satisfy}

In this Section, we take a phenomenological 
approach and consider a general kinetic function of the form:
\begin{equation}
\label{genaction}
P= -f_1 (\phi ) \sqrt{1-f_2 (\phi ) X} -f_3 (\phi) \,,
\end{equation}
where $f_i (\phi )$ are unspecified functions of the inflaton 
field\footnote{It is assumed 
implicitly that the functions have a suitable form for 
generating a successful phase of inflation.}.
A direct comparison with Eq. (\ref{defP}) 
indicates that the standard DBI action exhibits two important properties. 
The first is the condition that $f_1 f_2 =2$. This implies that 
$c_sP_{,X} =1$ and greatly simplifies the form of Eq. (\ref{rtheory}). 
The second property is that the warp factor uniquely determines 
the kinetic structure of the action, i.e., $h^4 \propto f_1 \propto f_2^{-1}$.  
In view of this, it is interesting to consider whether
the gravitational wave constraints could be weakened by relaxing one 
or both of these conditions. 

Eq. (\ref{genaction}) 
can be transformed into a similar form to that of 
Eq. (\ref{defP}) through the field redefinition 
\begin{equation}
\label{defvarphi}
\varphi \equiv \int d \phi \sqrt{\frac{f_1f_2}{2}}  \,.
\end{equation}
This implies that the sound speed of fluctuations in 
the inflaton is given by 
\begin{equation}
\label{generalspeed}
c_s = \sqrt{1-f_2 X} = \frac{f_1f_2}{2} \frac{1}{P_{,X}}  \,,
\end{equation}
and the scalar perturbation amplitude by 
\begin{equation}
\label{genamp}
P_S^2 = \frac{1}{2\pi^2}\frac{H^4}{f_1f_2\dot{\phi}^2}  \,.
\end{equation}
However, the consistency equation (\ref{consistency}) and 
non-Gaussianity constraint (\ref{fnlcs}) remain unaltered 
for this more general class 
of models \cite{lidser2}. It 
follows, therefore, 
that the CMB normalization condition (\ref{observewarp}) 
generalizes to a constraint on the value of $f_1 (\phi_*)$:  
\begin{equation}
\label{observef1}
\left( \frac{f_1}{\MP^4} \right)_{*} \simeq \frac{\pi^2}{16} r^2P_S^2
\left( 1+ \frac{1}{3f_{NL}} \right)  \,.
\end{equation}
Finally, the expression for the scalar spectral index
follows most directly by generalizing Eq. (\ref{obscon1}) 
via the correspondence (\ref{defvarphi}). It  
is straightforward to show that 
\begin{equation}
\label{constraintW}
1-n_s = \frac{r}{4} \sqrt{1+3f_{NL}}
 +\frac{n_{NL}}{1+3f_{NL} } \mp \sqrt{\frac{r}{4f_1f_2}} \left( 
\frac{f_1'}{f_1} \MP \right)_*  \,.
\end{equation}

Eq. (\ref{defvarphi}) also allows us to deduce 
that the infinitesimal variation of the effective 
field $\varphi$ is given by the Lyth bound, 
$(\Delta \varphi /\MP )^2 \simeq 
r (\Delta {\cal{N}})^2/8$. This implies that 
the variation of the inflaton field during inflation is 
\begin{equation}
\label{newbit}
\int_0^{\phi_{*}} \frac{d\phi}{\MP} \, \sqrt{\frac{f_1f_2}{2}} 
= \sqrt{\frac{r_*}{8}}  {\cal{N}}_{\mathrm{eff}} \,.
\end{equation}
If we now restrict our attention to 
the observable stage of inflation, and further assume that the variation
of $f_1f_2$ is negligible during that epoch, we find that 
\begin{equation}
\label{generalphivary1}
\left( \frac{\Delta \phi}{\MP} \right)^2_{*} \simeq 
\frac{(\Delta {\cal{N}}_{*} )^2}{4 f_1f_2} r_*  \,.
\end{equation}
On the other hand, the BM bound restricts the maximal 
variation of the scalar field $\phi$ in the throat region. 
This will be determined by expression (\ref{BMbound}) 
for generic warped geometries that are asymptotically 
$AdS_5 \times X_5$ away from the tip of the throat. Moreover, 
if observable scales leave the horizon 
while the brane is inside the throat, the change in the field value 
must satisfy $| \Delta \phi_*|<\phi_{UV}$. Hence, it follows from   
Eq. (\ref{generalphivary1}) that\footnote{We are 
being conservative by restricting our discussion to the 
observable phase of inflation. More generally, if 
$f_1f_2$ remains nearly constant over the last ${\cal{N}}$ 
e-foldings of inflation, 
we may substitute $\Delta {\cal{N}}_* \rightarrow 
{\cal{N}}_{\mathrm{eff}}$, where ${\cal{N}}_{\mathrm{eff}}$ 
may be as large as 60. Thus, the generalized bound 
(\ref{generalBMbound}) should be regarded as a necessary 
(but not sufficient) condition to be satisfied by the tensor modes.} 
\begin{equation}
\label{generalBMbound}
r_*< \frac{16}{N (\Delta {\cal{N}}_{*})^2} f_1f_2  \,.
\end{equation}

We will refer to condition 
(\ref{generalBMbound}) as the 
generalized BM bound. 
Combining expressions (\ref{observef1}) and (\ref{generalBMbound}) then
results in a necessary condition on the value of $f_2(\phi_*)$ 
for the generalized BM bound to be satisfied: 
\begin{equation}
\label{f2bound}
\frac{f_2 (\phi_*)\MP^4}{N} > \frac{(\Delta {\cal{N}}_*)^2}{\pi^2} 
\frac{1}{rP_S^2} \,.
\end{equation}
This condition can also be interpreted as a lower limit on $r$ 
for a given value of $f_2 (\phi_*)$.  

In UV models, identical arguments that led to 
the lower limit (\ref{lowerbound}) on the tensor-scalar ratio
will also apply in this more general context if, as expected, $f_1 (\phi) $ 
is a monotonically increasing function. A necessary condition for
the lower and upper limits
(\ref{lowerbound}) and (\ref{generalBMbound}) to be compatible, therefore, is  
that  
\begin{equation}
\label{evade}
f_1f_2 > \frac{N (\Delta {\cal{N}}_*)^2 (1-n_s)}{4\sqrt{3f_{NL}}} \,.
\end{equation}

In IR scenarios, however, the positive sign will apply in the 
last term of the right-hand side of Eq. (\ref{constraintW}).
Hence, assuming $f'_1 >0$ and neglecting the term proportional to 
$n_{NL}/f_{NL}$ yields an {\em upper} limit on the tensor-scalar ratio:
\begin{equation}
\label{IRupper}
r_* < \frac{4(1-n_s)}{\sqrt{3f_{NL}}}  \,.
\end{equation}
Combining conditions (\ref{f2bound}) and (\ref{IRupper}) 
therefore leads to a constraint on $f_2 (\phi_*) $
for the generalized BM bound to be satisfied in IR inflation: 
\begin{equation}
\label{f2IRlolwer}
\frac{f_2\MP^4}{N} > \frac{\sqrt{3} (\Delta {\cal{N}}_*)^2}{4\pi^2}
\frac{\sqrt{f_{NL}}}{(1-n_s)P_S^2}  \,.
\end{equation}

To summarize this Section, 
expression (\ref{generalBMbound}) implies that the 
BM bound (\ref{BMboundr}) 
could be relaxed if $f_1f_2 \gg 1$ on observable scales. 
It is therefore important to develop string-inspired models 
where this condition arises naturally. This will be the focus of the 
next Section.

\section{IR Inflation and Multiple Branes}

\label{sec:multiple}

Recently, an interesting version of IR DBI inflation 
was proposed by Thomas and Ward \cite{thomasward}. In this model, the flux annihilation process 
generates $n$ coincident branes that are initially located at the 
bottom of a throat region. The dynamics of this configuration
is determined by the non-Abelian world-volume theory \cite{myers1,myers2}. 
This theory exhibits extra stringy degrees of freedom which arise due to the 
fuzzy nature of the geometry. For the case where a fuzzy two-sphere is 
embedded in a three-cycle in the $X_5$ manifold, 
the kinetic structure of the action is given in the large $n$ limit by 
\cite{thomasward}
\begin{equation}
P=-nT_3 \left[ h^4(\phi ) W(\phi ) 
\sqrt{1-2 T_3^{-1} h^{-4}(\phi) X}
-h^4(\phi ) + V (\phi ) \right] \,,
\end{equation}
where   
\begin{equation} 
\label{defW}
W (\phi ) \equiv \sqrt{1+ C^{-1}h^{-4}(\phi ) \phi^4}
\end{equation}
defines the so-called `fuzzy' potential, 
$C = \pi^2 \hat{C}T_3^2/m_s^4$ is a model-dependent constant and 
$\hat{C} \simeq n^2$ 
is the quadratic Casimir of the $n$-dimensional representation of 
${\rm SU}(2)$. 
Comparison with Eq. (\ref{genaction}) 
implies that $f_1f_2 =2nW$ and $f_2=2/(T_3h^4)$. Hence, 
the new features of this model relative to the standard single-brane 
scenario are parametrized in terms of the fuzzy potential $W (\phi )$. 
This configuration is conjectured to be dual to 
a ${\rm D5}$-brane which is wrapped around a two-cycle 
of the throat \cite{dual1,dual2,dual3}. 

The regime $W \gg 1$ is of interest for 
relaxing the gravitational wave constraints\footnote{Note that 
the case $n \gg 1$ and
$W \sim 1$ will not significantly relax the BM bound, 
since we require $n \ll N$ for backreaction effects to be negligible.}. 
The generalized BM bound (\ref{f2bound}) may now be expressed as 
a limit on the value of the warp factor $h(\phi_*)$ on CMB scales: 
\begin{equation}
\label{warptorelax}
\frac{NT_3h^4_*}{\MP^4} < 
\frac{8\pi^2 (1-n_s)P_S^2}{\sqrt{3f_{NL}}(\Delta {\cal{N}}_*)^2} \,.
\end{equation}

We now consider whether this limit can be satisfied for reasonable choices 
of parameters when the warped compactification corresponds to 
an $AdS_5$ or KS throat, respectively. Recall that the warp 
factor for the $AdS_5$ throat is given by $h=\phi/(\sqrt{T_3}L)$.  
Condition (\ref{warptorelax}) therefore reduces to a constraint on the 
value of the inflation during observable inflation: 
\begin{equation}
\label{phivalue}
\frac{\phi_*^4}{\MP^4} < 
\frac{8\pi^2 (1-n_s)P_S^2}{\sqrt{3f_{NL}} ( \Delta {\cal{N}}_*)^2} 
\frac{\lambda}{N} \,,
\end{equation} 
where $\lambda \equiv \pi N/[2 \mathrm{Vol}(X_5)]$. 
However, non-perturbative string effects are expected to become 
important below a cutoff scale, $\phi_{\rm cut} = 
h_{\rm cut} \lambda^{1/4} m_s$, where $h_{\rm cut}$ is the value of the 
warp factor at that scale. For consistency, therefore, one requires 
$\phi_*>\phi_{\rm cut}$, which implies an upper limit on the 
${\rm D3}$-brane charge: 
\begin{equation}
\label{Nlimit}
N< \frac{8\pi^2 (1-n_s)P_S^2}{\sqrt{3f_{NL}} ( \Delta {\cal{N}}_*)^2}
\left( \frac{\MP}{h_{\rm cut}m_s} \right)^4  \,.
\end{equation}
Assuming the typical values $m_s \sim 10^{-2}\MP$, 
$\Delta {\cal{N}}_* \simeq 4$ and 
$h_{\rm cut} \sim 10^{-2}$ implies  
$N < 7 \times 10^7 (1-n_s)f_{NL}^{-1/2} < 2\times 10^6$, where 
the latter inequality follows for $1-n_s <0.05$ and $f_{NL} >5$. 

For an $AdS_5$ throat, the fuzzy potential 
is a constant and the condition that $W \gg 1$ becomes 
\begin{equation}
\label{Chatlimit}
\hat{C} \ll \frac{4\pi^2g_sN}{\mathrm{Vol}(X_5)} \,.
\end{equation}
Hence, combining inequalities 
(\ref{Nlimit}) and (\ref{Chatlimit}) implies that 
\begin{equation}
\label{nlimit}
\hat{C} \ll 
\frac{32 \pi^4 (1-n_s)P_S^2}{\sqrt{3f_{NL}} ( \Delta {\cal{N}}_*)^2}
\frac{g_s}{\mathrm{Vol} (X_5) }
\left( \frac{\MP}{h_{\rm cut}m_s} \right)^4  \,,
\end{equation}
and specifying $g_s \sim 10^{-2}$ and 
$\mathrm{Vol}(X_5) \simeq \pi^3$ then yields the limit  
$\hat{C} \ll 10^{6} (1-n_s)f_{NL}^{-1/2} < 2 \times 10^4$, or equivalently,  
$n \ll 150$. Furthermore, since $f_1f_2 \simeq {\rm constant}$, 
the inequality (\ref{generalBMbound}) may be strengthened by a 
factor of $(   {\cal{N}}_{\mathrm{eff}} /\Delta {\cal{N}}_*)^2$ by 
substituting 
$\Delta {\cal{N}}_* \rightarrow {\cal{N}}_{\mathrm{eff}}$. This ratio 
could be as 
high as $(60/4)^2 \simeq 200 $, which would rule out this particular
model. 

Since the branes are initially located at the tip of the 
throat, another case of interest is the IR limit of the KS geometry, where 
the warp factor asymptotes to a constant value 
\cite{gkp}:
\begin{equation}
\label{KStip}
h_{\rm tip} = \exp \left( - \frac{2\pi K}{3Mg_s} \right) \,,
\end{equation}
and $K,M \in {\rm \bf Z^+}$ are the units of flux associated 
with the NS-NS and R-R three-forms, respectively, such that $N=MK$.
In this case, the generalized BM bound (\ref{warptorelax}) becomes 
\begin{equation}
\label{loglimit}
\frac{8\pi K}{3Mg_s} -\ln N > 4 \ln \left( \frac{m_s}{g_s^{1/4}\MP} \right)
-\ln \left( 
\frac{64\pi^5 (1-n_s)P_S^2}{\sqrt{3f_{NL}} ( \Delta {\cal{N}}_*)^2}
\right) \,.
\end{equation}
The radius of the three-sphere at the tip of the KS 
throat is of the order $(g_sM)^{1/2}$ in string units 
and this must be large (and at the very least should exceed
unity) for the supergravity approximation to be reliable. 
Substituting this requirement into expression (\ref{loglimit}) 
results in a necessary (but not sufficient) condition 
on the ${\rm D3}$-brane charge for the 
generalized BM bound to be satisfied: 
\begin{equation}
\label{Nlowerlimit}
\frac{1}{N} \exp \left( \frac{8\pi g_s N}{3}  \right)
> \frac{\sqrt{3f_{NL}} ( \Delta {\cal{N}}_*)^2}{64\pi^5 
(1-n_s)P_S^2} \frac{1}{g_s} \left( \frac{m_s}{\MP} \right)^4 \,.
\end{equation}

Recalling that a necessary condition for the backreaction 
of the branes to be negligible is 
$N \gg n \gg 1$ implies that the 
exponential term in (\ref{Nlowerlimit}) will dominate unless 
the string coupling constant is extremely small. Hence, for 
the parameter estimations quoted above, we 
deduce the lower limit 
\begin{equation}
\label{yetanotherconstraint}
N- 12 \ln N > -6.8 +12 \ln \left( \frac{\sqrt{f_{NL}}}{1-n_s} \right) \,,
\end{equation}
which becomes $N \, \gsim \, 10^2$ for $1-n_s \simeq 0.05$ and $f_{NL}>5$.

In general, however, the $K$ and $M$ units of flux are not independent. 
F-theory compactification on Calabi-Yau four-folds
provides a geometric way of parameterizing  
type IIB string compactifications
\cite{witten1,witten2,witten3,sethi,gkp,klemm}. 
Global tadpole cancellation constrains the topology of the four-fold
and this restricts the brane and flux configurations.  
When the KS system is embedded into F-theory, the  
constraint is given by \cite{gkp}
\begin{equation}
\label{Ftheory}
\frac{\chi}{24} = n + MK \,,
\end{equation}
where $\chi$ is the Euler characteristic of the four-fold.  
Hence, $N = MK < \chi /24$ and together with condition 
(\ref{yetanotherconstraint}), this implies that
\begin{equation}
\label{chilimit}
\chi > 2400 \,,
\end{equation}
for $N > 10^2$.
It is known that the Euler number for four-folds 
corresponding to hypersurfaces in weighted projective spaces
can be as high as $\chi \le 1,820,448$ \cite{klemm},
so there are many compactifications that could 
in principle satisfy the generalized BM bound.
On the other hand, the above limit on the Euler characteristic 
does allow us to gain some insight into the 
topology of the extra dimensions, since compactifications 
which result in a small Euler characteristic would be  
incompatible with the generalized BM bound. 
 
\section{Conclusion} 

\label{Conclusion}

In this paper, we have derived an upper limit to
the amplitude of the primordial gravitational wave spectrum
generated during UV DBI inflation. We considered   
the maximal inflaton field variation   
that can occur during the observable stages of inflation and assumed  
only that the brane was inside the throat during that epoch. 
The bound (\ref{upperbound}) is valid for an arbitrary inflaton potential and 
warp factor (modulo some weak caveats) and can be expressed 
entirely in terms of observable parameters once the volume of 
the five-dimensional sub-manifold of the throat has been specified. 
The inferred upper limit on $r$ is surprisingly strong. 
We find that the standard UV  
scenario predicts tensor perturbations that are undetectably small, 
at a level ${r_*} \, {\lsim} \, {10^{-7}}$. 

The current WMAP3 data 
favours models that generate a red spectral index $n_s<1$
when both the gravitational waves and running in the scalar 
spectral index are negligible. For UV versions of the scenario, 
we have identified a corresponding 
lower limit on $r$ which applies in this region of 
parameter space, $r_* \, \gsim \, 0.1 (1-n_s)$. It is clear that 
the standard model 
can not satisfy both the upper and lower bounds 
on the tensor modes for the observationally favoured value 
$1-n_s \simeq 0.05$.

The generality of our 
analysis implies that modifying either the inflaton potential 
or the form of the warp factor is unlikely to resolve this discrepancy. 
On the other hand, there are a number of possible ways of reconciling  
theory with observation. In general, 
either the upper or lower limit on $r$ needs to be relaxed. 
Weakening the latter would require a violation of the slow-roll 
conditions or a blue spectral index. 
A value of $n_s >1$ is compatible with WMAP3 if the running of the 
spectral index 
is sufficiently negative, but is only marginally
consistent if just the tensor modes are non-negligible.  The 
upper limit on $r$ can be weakened by reducing 
the volume of $X_5$ or 
by generalizing the DBI action. Furthermore, it need not necessarily 
apply in IR versions of the scenario, although the BM bound will still hold
in such cases. 

We considered a generalized version of the 
DBI action and identified a necessary condition on the form of such  
an action for the BM bound to be relaxed.
As a concrete example, 
we investigated a version of IR inflation that is driven by 
multiple coincident branes and found that  
the bounds on the tensor-scalar ratio can indeed 
be made compatible if the brane charges satisfy appropriate 
conditions.   

The upper bound on $r$ that we derived in section 3 arises 
because the warp factor in standard DBI models 
completely specifies the kinetic 
energy of the inflaton field. Deriving a corresponding bound for 
the generalized model (\ref{genaction}) would be more involved, 
since the CMB normalization (\ref{observef1}) only 
directly constrains the function 
$f_1 (\phi )$ and this may not necessarily depend on the warp factor. 
Nonetheless, the constraints (\ref{halfwayconstraint}) and
(\ref{trapeziumlimit}) can be 
combined with Eq. (\ref{generalphivary1}) to derive a limit 
on the tensor-scalar ratio in terms of the warp factor and the 
kinetic function (\ref{genaction}) and we find that  
\begin{equation}
\label{generalgeneral}
r_* < \frac{2}{\pi^{2/3} ( \Delta \mathcal{N}_*)^2}
\left( \frac{m_s}{\MP} \right)^{4/3} 
\frac{h_*^{4/3} (f_1f_2)_*}{g_s\mathrm{Vol}(X_5)}  \,.
\end{equation}
For a specific model where the warp factor and 
the functions $f_i (\phi )$ are determined by particle 
physics considerations,   
condition (\ref{generalgeneral}) 
may be interpreted as a bound that relates 
the tensor modes directly to the value of the inflaton field during observable 
inflation. This constraint provides a consistency 
check that any given model must satisfy 
irrespective of the form of the inflaton potential. 

In conclusion, therefore, primordial gravitational wave constraints 
combined with cosmological observations of the density perturbation
spectrum act as a powerful discriminant of DBI inflationary models. 
They also serve as an important 
observational guide for identifying viable 
generalizations of the scenario. 

\ack
IH is supported jointly by a Queen Mary studentship and 
the Science and Technology Facilities Council (STFC). 
We thank S. Thomas and J. Ward for helpful discussions. 

\section*{References}



\begin{thebibliography}{10}
\expandafter\ifx\csname url\endcsname\relax
  \def\url#1{{\tt #1}}\fi
\expandafter\ifx\csname urlprefix\endcsname\relax\def\urlprefix{URL }\fi
\providecommand{\eprint}[2][]{\url{#2}}

\bibitem{spergel}
Spergel D~N {\em et~al.\/} (WMAP), {\em {W}ilkinson {M}icrowave {A}nisotropy
  {P}robe ({WMAP}) three year results: {I}mplications for cosmology\/}, 2006
  [\eprint{astro-ph/0603449}]

\bibitem{brane1}
Dvali G~R and Tye S~H~H, {\em Brane inflation\/}, 1999 {\em Phys. Lett.\/} {\bf
  B450} 72--82 [\eprint{hep-ph/9812483}]

\bibitem{brane2}
Dvali G~R, Shafi Q and Solganik S, {\em {D}-brane inflation\/}, 2001
  [\eprint{hep-th/0105203}]

\bibitem{brane3}
Burgess C~P, Majumdar M, Nolte D, Quevedo F, Rajesh G and Zhang R~J, {\em The
  inflationary brane-antibrane universe\/}, 2001 {\em JHEP\/} {\bf 07} 047
  [\eprint{hep-th/0105204}]

\bibitem{brane4}
Kachru S, Kallosh R, Linde A and Trivedi S~P, {\em {D}e {S}itter vacua in
  string theory\/}, 2003 {\em Phys. Rev.\/} {\bf D68} 046005
  [\eprint{hep-th/0301240}]

\bibitem{brane5}
Kachru S, Kallosh R, Linde A, Maldacena J, McAllister L and Trivedi S~P, {\em
  Towards inflation in string theory\/}, 2003 {\em JCAP\/} {\bf 0310} 013
  [\eprint{hep-th/0308055}]

\bibitem{brane6}
Silverstein E and Tong D, {\em Scalar speed limits and cosmology:
  {A}cceleration from {D-}cceleration\/}, 2004 {\em Phys. Rev.\/} {\bf D70}
  103505 [\eprint{hep-th/0310221}]

\bibitem{brane7}
Shandera S, Shlaer B, Stoica H and Tye S~H~H, {\em Inter-brane interactions in
  compact spaces and brane inflation\/}, 2004 {\em JCAP\/} {\bf 0402} 013
  [\eprint{hep-th/0311207}]

\bibitem{brane8}
Firouzjahi H and Tye S~H~H, {\em Closer towards inflation in string theory\/},
  2004 {\em Phys. Lett.\/} {\bf B584} 147--154 [\eprint{hep-th/0312020}]

\bibitem{brane9}
Burgess C~P, Cline J~M, Stoica H and Quevedo F, {\em Inflation in realistic
  {D}-brane models\/}, 2004 {\em JHEP\/} {\bf 09} 033 [\eprint{hep-th/0403119}]

\bibitem{brane10}
Iizuka N and Trivedi S~P, {\em An inflationary model in string theory\/}, 2004
  {\em Phys. Rev.\/} {\bf D70} 043519 [\eprint{hep-th/0403203}]

\bibitem{brane11}
Alishahiha M, Silverstein E and Tong D, {\em {DBI} in the sky\/}, 2004 {\em
  Phys. Rev.\/} {\bf D70} 123505 [\eprint{hep-th/0404084}]

\bibitem{brane12}
Chen X, {\em Multi-throat brane inflation\/}, 2005 {\em Phys. Rev.\/} {\bf D71}
  063506 [\eprint{hep-th/0408084}]

\bibitem{brane13}
Chen X, {\em Inflation from warped space\/}, 2005 {\em JHEP\/} {\bf 08} 045
  [\eprint{hep-th/0501184}]

\bibitem{brane14}
Chen X, {\em Running non-{G}aussianities in {DBI} inflation\/}, 2005 {\em Phys.
  Rev.\/} {\bf D72} 123518 [\eprint{astro-ph/0507053}]

\bibitem{brane15}
Cline J~M and Stoica H, {\em Multibrane inflation and dynamical flattening of
  the inflaton potential\/}, 2005 {\em Phys. Rev.\/} {\bf D72} 126004
  [\eprint{hep-th/0508029}]

\bibitem{brane16}
Frey A~R, Mazumdar A and Myers R, {\em Stringy effects during inflation and
  reheating\/}, 2006 {\em Phys. Rev.\/} {\bf D73} 026003
  [\eprint{hep-th/0508139}]

\bibitem{brane17}
Chialva D, Shiu G and Underwood B, {\em Warped reheating in multi-throat brane
  inflation\/}, 2006 {\em JHEP\/} {\bf 01} 014 [\eprint{hep-th/0508229}]

\bibitem{Brodie:2003qv}
Brodie J~H and Easson D~A, {\em Brane inflation and reheating\/}, 2003 {\em
  JCAP\/} {\bf 0312} 004 [\eprint{hep-th/0301138}]

\bibitem{Vikman:2006hk}
Vikman A, {\em Inflation with large gravitational waves\/}, 2006
  [\eprint{astro-ph/0606033}]

\bibitem{Mukhanov:2005bu}
Mukhanov V~F and Vikman A, {\em Enhancing the tensor-to-scalar ratio in simple
  inflation\/}, 2006 {\em JCAP\/} {\bf 0602} 004 [\eprint{astro-ph/0512066}]

\bibitem{Kallosh:2007wm}
Kallosh R and Linde A, {\em Testing {S}tring {T}heory with {CMB}\/}, 2007 {\em
  JCAP\/} {\bf 0704} 017 [\eprint{arXiv:0704.0647 [hep-th]}]

\bibitem{brane18}
Shandera S~E and Tye S~H~H, {\em Observing brane inflation\/}, 2006 {\em
  JCAP\/} {\bf 0605} 007 [\eprint{hep-th/0601099}]

\bibitem{brane19}
Chen X and Tye S~H~H, {\em Heating in brane inflation and hidden dark
  matter\/}, 2006 {\em JCAP\/} {\bf 0606} 011 [\eprint{hep-th/0602136}]

\bibitem{brane20}
Kecskemeti S, Maiden J, Shiu G and Underwood B, {\em {DBI} inflation in the tip
  region of a warped throat\/}, 2006 {\em JHEP\/} {\bf 09} 076
  [\eprint{hep-th/0605189}]

\bibitem{brane21}
Shiu G and Underwood B, {\em Observing the geometry of warped compactification
  via cosmic inflation\/}, 2007 {\em Phys. Rev. Lett.\/} {\bf 98} 051301
  [\eprint{hep-th/0610151}]

\bibitem{tyereview}
Tye S~H, {\em Brane inflation: {S}tring theory viewed from the cosmos\/}, 2006
   [\eprint{hep-th/0610221}]

\bibitem{cline}
Cline J~M, {\em String cosmology\/}, 2006   [\eprint{hep-th/0612129}]

\bibitem{lidser3}
Seery D and Lidsey J~E, {\em Primordial non-gaussianities in single field
  inflation\/}, 2005 {\em JCAP\/} {\bf 0506} 003 [\eprint{astro-ph/0503692}]

\bibitem{chenetal}
{Chen} X, {Huang} {\ M-x}, {Kachru} S and {Shiu} G, {\em Observational
  signatures and non-{G}aussianities of general single field inflation\/}, 2007
  {\em JCAP\/} {\bf 0701} 002 [\eprint{hep-th/0605045}]

\bibitem{bmpaper}
Baumann D and McAllister L, {\em A microscopic limit on gravitational waves
  from {D}-brane inflation\/}, 2006   [\eprint{hep-th/0610285}]

\bibitem{lyth}
Lyth D~H, {\em What would we learn by detecting a gravitational wave signal in
  the cosmic microwave background anisotropy?\/}, 1997 {\em Phys. Rev. Lett.\/}
  {\bf 78} 1861--1863 [\eprint{hep-ph/9606387}]

\bibitem{bean}
Bean R, Shandera S~E, Henry~Tye S~H and Xu J, {\em Comparing brane inflation to
  {WMAP}\/}, 2007 {\em JCAP\/} {\bf 0705} 004 [\eprint{hep-th/0702107}]

\bibitem{clover}
Taylor A~C {\em et~al.\/}, {\em Clover: A new instrument for measuring the
  {B}-mode polarization of the {CMB}\/}, 2004   [\eprint{astro-ph/0407148}]

\bibitem{vpj}
Verde L, Peiris H and Jimenez R, {\em Optimizing {CMB} polarization experiments
  to constrain inflationary physics\/}, 2006 {\em JCAP\/} {\bf 0601} 019
  [\eprint{astro-ph/0506036}]

\bibitem{thomasward}
Thomas S and Ward J, {\em {IR} inflation from multiple branes\/}, 2007
  [\eprint{hep-th/0702229}]

\bibitem{grana}
Grana M, {\em Flux compactifications in string theory: {A} comprehensive
  review\/}, 2006 {\em Phys. Rept.\/} {\bf 423} 91--158
  [\eprint{hep-th/0509003}]

\bibitem{douglas}
Douglas M~R and Kachru S, {\em Flux compactification\/}, 2006
  [\eprint{hep-th/0610102}]

\bibitem{ks}
Klebanov I~R and Strassler M~J, {\em Supergravity and a confining gauge theory:
  Duality cascades and $\chi${SB}-resolution of naked singularities\/}, 2000
  {\em JHEP\/} {\bf 08} 052 [\eprint{hep-th/0007191}]

\bibitem{kt}
Klebanov I~R and Tseytlin A~A, {\em Gravity duals of supersymmetric ${SU(N)}
  \times {SU(N+M)}$ gauge theories\/}, 2000 {\em Nucl. Phys.\/} {\bf B578}
  123--138 [\eprint{hep-th/0002159}]

\bibitem{gm}
Garriga J and Mukhanov V~F, {\em Perturbations in k-inflation\/}, 1999 {\em
  Phys. Lett.\/} {\bf B458} 219--225 [\eprint{hep-th/9904176}]

\bibitem{lidser1}
Lidsey J~E and Seery D, {\em An observational test of holographic inflation\/},
  2006 {\em Phys. Rev.\/} {\bf D73} 023516 [\eprint{astro-ph/0511160}]

\bibitem{lidser2}
Lidsey J~E and Seery D, {\em Primordial non-{G}aussianity and gravitational
  waves: {O}bservational tests of brane inflation in string theory\/}, 2007
  {\em Phys. Rev.\/} {\bf D75} 043505 [\eprint{astro-ph/0610398}]

\bibitem{maldacena}
Maldacena J~M, {\em Non-{G}aussian features of primordial fluctuations in
  single field inflationary models\/}, 2003 {\em JHEP\/} {\bf 05} 013
  [\eprint{astro-ph/0210603}]

\bibitem{crim}
Creminelli P, Senatore L, Zaldarriaga M and Tegmark M, {\em Limits on $f_{NL}$
  parameters from {WMAP} 3yr data\/}, 2007 {\em JCAP\/} {\bf 0703} 005
  [\eprint{astro-ph/0610600}]

\bibitem{planck}
Planck project, \urlprefix \url{http://www.rssd.esa.int/index.php?project=Planck}

\bibitem{gauntlett}
Gauntlett J~P, Martelli D, Sparks J and Waldram D, {\em {S}asaki-{E}instein
  metrics on {S(2) x S(3)}\/}, 2004 {\em Adv. Theor. Math. Phys.\/} {\bf 8}
  711--734 [\eprint{hep-th/0403002}]

\bibitem{myers1}
Myers R~C, {\em Dielectric-branes\/}, 1999 {\em JHEP\/} {\bf 9912} 022
  [\eprint{hep-th/9910053}]

\bibitem{myers2}
Myers R~C, {\em Nonabelian phenomena on {D}-branes\/}, 2003 {\em Class. Quantum
  Grav.\/} {\bf 20} S347 [\eprint{hep-th/0303072}]

\bibitem{dual1}
Thomas S and Ward J, {\em Non-{A}belian (p,q) strings in the warped deformed
  conifold\/}, 2006 {\em JHEP\/} {\bf 12} 057 [\eprint{hep-th/0605099}]

\bibitem{dual2}
Firouzjahi H, {\em Dielectric (p,q) strings in a throat\/}, 2006 {\em JHEP\/}
  {\bf 12} 031 [\eprint{hep-th/0610130}]

\bibitem{dual3}
Bachas C, Douglas M~R and Schweigert C, {\em Flux stabilization of
  {D}-branes\/}, 2000 {\em JHEP\/} {\bf 05} 048 [\eprint{hep-th/0003037}]

\bibitem{gkp}
Giddings S~B, Kachru S and Polchinski J, {\em Hierarchies from fluxes in string
  compactifications\/}, 2002 {\em Phys. Rev.\/} {\bf D66} 106006
  [\eprint{hep-th/0105097}]

\bibitem{witten1}
Witten E, {\em Phase transitions in {M}-theory and {F}-theory\/}, 1996 {\em
  Nucl. Phys.\/} {\bf B471} 195 [\eprint{hep-th/9603150}]

\bibitem{witten2}
Sethi S, Vafa C and Witten E, {\em Constraints on low-dimensional string
  compactifications\/}, 1996 {\em Nucl. Phys.\/} {\bf B480} 213
  [\eprint{hep-th/9606122}]

\bibitem{witten3}
Gukov S, Vafa C and Witten E, {\em {CFT}'s from {C}alabi-{Y}au four-folds\/},
  2000 {\em Nucl. Phys.\/} {\bf B584} 69 [\eprint{hep-th/9906070}]

\bibitem{sethi}
Dasgupta K, Rajesh G and Sethi S, {\em M-theory, orientifolds and {G}-flux\/},
  1999 {\em JHEP\/} {\bf 9908} 023 [\eprint{hep-th/9908088}]

\bibitem{klemm}
Klemm A, Lian B, Roan S~S and Yau S~T, {\em Calabi-{Y}au fourfolds for {M}- and
  {F}-theory compactifications\/}, 1998 {\em Nucl. Phys.\/} {\bf B518} 515
  [\eprint{hep-th/9701023}]

\end{thebibliography}

\providecommand{\newblock}{}

\end{document}